\documentclass[12pt,preprint]{aastex}

\shorttitle{Stability analysis  of  the  HD 82943 and HD 37124
planetary systems } \shortauthors{Ji Jianghui et al.}

\slugcomment{Submitted to \textit{The Astrophysical Journal}}

\begin{document}

\title{The  Stability Analysis  of  the  HD 82943 and HD 37124 Planetary  Systems}
\author{JI  Jianghui\altaffilmark{1,3}, Kinoshita Hiroshi\altaffilmark{4}, LIU
Lin\altaffilmark{2,3}, LI Guangyu\altaffilmark{1,3}, Nakai
Hiroshi\altaffilmark{4}} \email{jijh@pmo.ac.cn }

\altaffiltext{1}{Purple  Mountain  Observatory , Chinese  Academy
of  Sciences ,  Nanjing  210008,China}
\altaffiltext{2}{Department
of Astronomy,  Nanjing University, Nanjing  210093, China}

\altaffiltext{3}{National Astronomical Observatory, Chinese
Academy of Sciences,Beijing 100012,China}
\altaffiltext{4}{National Astronomical Observatory,
 Mitaka, Tokyo 181-8588,Japan}

\begin{abstract}

We carry out numerical simulations to explore the dynamical
evolution of the HD 82943 and HD 37124 planetary systems,which
both have two Jupiter-like planets. By simulating various
planetary configurations in the neighborhood of the fitting
orbits, we find three mechanisms to maintain the stability of
these systems: For HD 82943,we find that the 2:1 mean motion
resonance can act as the first mechanism  for  all the stable
orbits. The second mechanism is the alignment of the periastron of
the two planets of  HD 82943 system. In the paper,we show one case
is simultaneously maintained by the two mechanisms.
Additionally,we also use the corresponding analytical models
successfully to explain the different numerical results for the
system. The third mechanism is the Kozai resonance which takes
place in the mutual highly orbits of HD 37124. In the
simulations,we discover that the argument of periastron $\omega$
of the inner planet librates about $90^{\circ}$ or $270^{\circ}$
for the whole time span. The Kozai mechanism can explain the
stable configuration of large eccentricity of the inner planet.
\end{abstract}


\section{Introduction}
At present, 88 planetary systems,\footnote{See
http://www.obspm.fr/planets}containing 101 giant extrasolar
planets,have been discovered in Doppler surveys of solar-type
stars(Marcy et al. 2000; Butler et al. 2000,2001,2002), among
which there are 11 multiplanetary systems,including 9 two-planet
systems(HD 83443, HD 82943, GJ 876, HD 168443, HD 74156, 47 Uma,
HD 37124, HD 38529 and HD 12661) and 2 three-planet systems(55 Cnc
and Ups And). In this paper,we are focusing on two planetary
systems of HD 82943 and HD 37124. An international team of
astronomers from the Geneva Observatory and other research
institutes announced the discovery of  HD 82943 planetary
system,which contains two Jupiter-like planets move around the
main star. And a new planet of  HD 37124 was recently found by the
scientists from the Keck Observatory, the University of California
and the California Institute of Technology.

HD 82943 is a  G0 star with $B-V=0.623$\footnote{See
http://obswww.unige.ch/$\sim$udry/planet/hd82943syst.html},
Hipparcos parallax of 36.42 mas and distance of 27.46 pc. The mass
of the parent star is $1.05 M_{\odot}$. Similar to the GJ 876
system(Marcy et al. 2001;Lee \& Peale 2002),which are suggested
that the stability of the planetary system might be sustained by
the 2:1 mean-motion resonance(MMR) and secular resonance,the two
planets of  HD 82943 are now also locked in the state of 2:1
resonance, with orbital periods 444.6\,d and 221.6\,d, and
semi-major axes 1.16\,AU\ and 0.73\,AU.  HD 37124 is a G4 dwarf
star with $B-V=0.667$(Butler 2002,see their Table 1). The mass of
the star is $0.91 M_{\odot}$. The orbital periods of  two
companions of  HD 37124  are 153\,d and 1942\,d repspectively,
with semi-major axes 0.54\,AU\ and 2.95\,AU.

In our earlier paper(Ji,Li \& Liu 2002), we performed simulations
to explore the dynamics of the  GJ 876 system by integrating
different planetary configurations of two-planet system. In this
paper,we aim to study the dynamical behaviors of the  HD 82943 and
HD 37124 systems by examining  various orbits in the vicinity of
the fitting results,further attempt to discover the possible
mechanisms to maintain these systems,and then make the theoretical
explanations of them.

In  the remainder of  this  paper is  organized as follows: In
Section 2,we introduce the numerical setup of the dynamical
simulations. In Section 3, we present the main outcomes of the
simulations of  the HD 82943 and HD 37124 systems and then
describe the analytical models that helps to explain stable
mechanisms of the two systems. In Section 4, we summarize the
chief results and give a brief discussion.

\section{Numerical setup}

On the basis of the N-body codes(Ji,Liu \& Li 2002), our numerical
integrations were performed for two double-planet systems.In our
simulations,the masses of the central star are separately  $1.05
M_{\odot}$ and $0.91 M_{\odot}$. By assuming sin\textit{i} = 1,we
obtained  the masses of four planets-HD82943b,HD82943c,HD37124b
and HD37124c are $1.63 M_{Jup}$, $0.88 M_{Jup}$, $0.86 M_{Jup}$
and $1.01 M_{Jup}$ respectively.  In the simulations,we adopted
time step of one percent of the orbital period of the inner planet
of each system when integrating two planets . The numerical errors
were effectively controlled during the integration,with the local
truncation error $10^{-14}$ for the time span of 10$^{7}$ yr,in
the same time the accuracy was also checked by the energy errors.
In comparison, we also used symplectic integrators (Feng
1986;Wisdom \& Holman 1991) to integrate the same orbit to assure
the results for some cases.

For HD 82943,we took the following steps to generate the initial
orbits to start the integration procedure :for each planet,we need
to obtain six orbital elements - semi-major axis $a$ ,
eccentricity $e$ , inclination $i$, nodal longitude $\Omega $,
periastron $\omega $ and mean anomaly $M$. At first, the two
planets are considered to locate in the same orbital plane. The
orbital parameters(see Table 1) are partly taken from the web
site.\footnote{See http://www.obspm.fr/encycl.html,
http://obswww.unige.ch/$\sim$udry/planet/hd82943syst.html} For all
the orbits, we assume that the semi-major axes of the two planets
are always unchanged,i.e.,1.16\,AU\ and 0.73\,AU respectively. And
their initial inclinations are taken as $0.5^{\circ}$ for all
cases. The eccentricities are constructed as follows:we let the
initial eccentricities $e_{0}$,0.41 and 0.54 respectively, be the
centers and take the error of the measurement $\Delta e$ as the
radii to randomly produce the eccentricities,such that $e
\in[0.41-0.08,0.41+0.08]$ and $[0.54-0.05,0.54+0.05]$
respectively. The same action can be done for the arguments of
periastra,then we separately obtain
$\omega\in[117.8^{\circ}-3.4^{\circ},117.8^{\circ}+3.4^{\circ}]$
and $[138.0^{\circ}-10.2^{\circ},138.0^{\circ}+10.2^{\circ}]$. The
remained two angles of nodal longitudes and mean anomalies are
randomly chosen between $0^{\circ}$ and $360^{\circ}$. Finally,we
obtained 100 orbits for each planet, which were  numbered from 1
to 100 in pairs, to perform the integration of the planetary
system,and each pair of the orbits was integrated for the time
span of 10 Myr.

The similar treatment can be done to yield the initial  orbital
elements of HD 37124, and  the uncertainties of the
eccentricities(Butler et al. 2002)and the arguments of the
periastron are together listed in Table 1. However, for HD
37124,we not only consider the coplanar orbits but also take into
account the mutual inclined orbits. Obviously,these orbits of the
two planetary systems lie in the neighborhood of the fitting
results,so we attempt to discover the stable configurations of
these systems.

\section{Dynamical analysis }
In this section,we will introduce the results of the dynamical
simulations of HD 82943 and HD 37124 planetary systems.
Additionally,we further study the possible mechanisms of
maintaining the two systems.

\subsection{HD 82943}

\subsubsection{Results of the dynamical simulations}
For  HD 82943, we totally completed 100 integrations for the time
span of 10 Myr. By cataloguing  the numerical results, we found
that 75 out of 100 coupled orbits are unstable for the time span
of $10^{5}$ yr. However, for a longer integration of 1 Myr, only 6
out of 25 cases are stable,15 out of 25  correspond to the cases
of the escape of the inner planet and 4 out of 25 are associated
with the leave of the outer planet. Finally, we obtained 5 pairs
of the orbits, numbered No.1,No.8,No.17, No.37 and No.92
respectively, which remain stable for 10 Myr, whereas the case of
No.86 is unstable when the system evolves at the time $T>2.63$
Myr. In the simulations, we notice that all the stable cases are
involved in the 2:1 MMR and easily understand that the stability
of a system is sensitive to its initial planetary configuration.
The simulation results show that most systems  tend to
self-destruct in 10$^{4}$-10$^{5}$ yr, and the lifetime is even
shorter for some cases. However, we are concentrating on the
stable cases.  In Table 2 are presented the ranges of the
variation of the semi-major axis \textit{a}, eccentricity
\textit{e} and inclination \textit{i} of the two planets over the
time span of 10 Myr for five stable configurations.From the
table,we  note that the eccentricity of the inner planet can be
rather high,even the maximum is $0.7392$ for one case ,while the
eccentricity of the outer planet appears to be moderate or in
nearly circular orbits,and we observe that the cases of the 2:1
MMR commonly occur in the highly inclined orbits during the
orbital evolution. In addition, Fig.1 displays the orbital
variations of the inner and outer planets for the case of No.17;
in  particular, we  see that the semi-major axes would undergo
small oscillations for the whole time span.

\subsubsection{2:1 mean motion resonance}
The resonances between the major planets (Murray \& Dermott
1999)are very common phenomena in our solar system.We know that
Jupiter and Saturn are in a 5:2 near-resonance,and the planets
Neptune and Pluto are in a peculiar 3:2 mean motion resonance that
maximize their separation to prevent them from close approaches
when they are at conjunction. The MMR frequently occurs for the
moons of the major planets,the asteroid belt and
recently-discovered Kuiper belt objects(Duncan et al. 1995;Wan \&
Huang 2001;Zhou et al. 2000,2002). Additionally, more and more
extrasolar planets of the multi-planetary systems are found in the
MMR,so our interest is to study them in these systems.

In the usual notation, the critical argument $\theta_{1}$,
$\theta_{2}$ for the 2:1 MMR is

\begin{equation}
\label{eq1} \quad\quad\theta_{1} = \lambda _{1} - 2\lambda _{2} +
\tilde {\omega} _{1},
\end{equation}
\begin{equation}
\label{eq2} \quad\quad\theta_{2} = \lambda _{1} - 2\lambda _{2} +
\tilde {\omega} _{2},
\end{equation}

\noindent Here $\lambda _{1} $, $\lambda _{2} $ are ,
respectively,the mean longitudes of the inner and outer planets,
and $\tilde {\omega} _{1} $, $\tilde {\omega} _{2}$ respectively
denote the periastra longitudes of the two planets(hereafter
subscript 1 denotes the inner planet, 2 the outer planet).

For the orbit of No. 8, Fig. 2a illustrates two resonant arguments
$\theta_{1}$ and $\theta_{2}$  vary with time for the time span of
10 Myr. From the figure,we notice that  $\theta_{1}$ librates
about $180^{\circ}$,while  $\theta_{2}$ librates about
$0^{\circ}$. Furthermore,Fig.2b in greater details displays the
semi-major axes of the inner and outer planets of the system
plotted against the resonant arguments,in which the semi-major
axes undergo small oscillations for the overall time span and the
equilibrium points are $(0.730,180^{\circ})$ and
$(1.165,0^{\circ})$ respectively. Sequentially,these two figures
exhibit that the two planets are really in the 2:1 MMR, then imply
that this kind of the resonance may be responsible for the
stability of the HD 82943 system.

The argument  $\theta_{1}$ librates around $0^{\circ}$ for the
orbits of No.1,No.17 and No.37, and the equilibrium point is
$(0.730,0^{\circ})$ for three cases. Because of the libration of
$\theta_{1}$, the semi-major axis of HD 82943c is well preserved
to prevent from larger changes. On the other hand,we also found
similar phenomena for the cases of HD 82943b. In Section 3.1.3,we
will discuss the corresponding mechanism by using the coplanar
model for the semi-major axis. However,we again discovered the
circulation of $\theta_{1}$ ranging from $0^{\circ}$ to
$360^{\circ}$ for No. 86,which may lead to the instability of the
system  and the planetary system de-integrates over 2.63 Myr.

\subsubsection{Coplanar analytical model for the semi-major axis}

From Table 2,we observe that for some stable cases the two planets
of the system can have low inclinations,and the eccentricity of
the outer planet can have small value and approach to 0 during the
orbital evolution.So we consider an averaged plane elliptical
model in the context of three-body system,by considering the
effect of the eccentricity of the outer planet. In the
computation,the units of the mass,the length and the time are
adopted as our previous paper.

Following our earlier paper (Ji, Liu \& Li 2002;Ji, Liu \& Liao
2000), we  have the conjugated action-angle variables(hereafter
subscript 1 is omitted for simplicity):

\begin{equation}
\label{eq3}
\left\{
\begin{array}{l}
 \tilde {L} = \sqrt{a}  \\
 \tilde {l} = \theta_{1}= \lambda-2\lambda_{2} + \tilde{\omega}  \\
\end{array}
\right.
\end{equation}

\noindent Then the corresponding Hamiltonian, by eliminating
short-periodic terms and ignoring high eccentricity terms,takes
the following form:

\begin{equation}
\label{eq4}
F = \tilde {F}_0 \left(\tilde {L}\right) + \tilde
{F}_c \left(\tilde {L},\mbox{ }\tilde {G}\right) + \tilde
{F}_{1}\left( \tilde {L},\tilde {l},\tilde {G};e_{2},\theta
\right)
\end{equation}

\noindent where

\begin{equation}
\label{eq5}
\left\{ {\begin{array}{l}
  \mbox{ }\tilde {F}_1 \mbox{ } =
 - \mu _2 a^2\left[ {\mbox{ }\left( {\frac{9}{4} + \frac{5}{4}a^2}
\right)e\cos \tilde {l} - \frac{9}{8}ae_2 \cos (\tilde {l} -
\theta )\mbox{
}} \right]\mbox{ ,} \\
\mbox{ } \theta = \mbox{ }\tilde {\omega } - \tilde {\omega }_2 =
\mbox{ }(\omega + \Omega ) -(\omega _2 + \Omega _2 )\mbox{ . } \\
\end{array}} \right.
\end{equation}

\noindent where in the equation(\ref{eq5}), $\mu _{2} $  and
$e_{2}$ respectively, are the mass and the eccentricity of HD
82943b. Clearly, $\theta$ is equal to $\theta_{3}$(see Section
3.1.4.). Lee \& Peale(2002) suggested that $\theta_{3}$ is
librating about $0^{\circ}$ for the GJ 876 planetary system as a
stable mechanism to maintain the system.In Section 3.1.4.,we will
show that $\theta_{3}$ librates near $180^{\circ}$ for the HD
82943 system. Therefore,for the second part (in brackets) of the
right-sided of $\tilde F_{1}$ in the equation (\ref{eq5}),we have
$|-\frac{9}{8}ae_{2}\cos(\tilde{l}-\theta)|\simeq|-\frac{9}{8}ae_{2}\cos\tilde{l}|$,
which can have the magnitude of the former item. Then there comes
up a question: how does this part influence on the motion of the
inner planet? We discuss this problem in the following.

However,as the Hamiltonian is not integrable,we should make some
assumptions to simplify (\ref{eq4}): firstly,we assume $\tilde
{G}=\tilde {G}_0$ and $\theta=\theta_{0}$(for example,$0^{\circ}$
or $180^{\circ}$) for the adiabatic variations,which indicate that
they change slowly with the time.Further we suppose $e_{2}=e_{20}$
in the computation. Finally, we have the Hamiltonian of a one
degree of freedom system:

\begin{equation}
\label{eq6}
 F = \tilde {F}_{0} \left( {\tilde {L}} \right) + \tilde {F}_{c} \left(
{\tilde {L} , \tilde {G}_{0}} \right) + \tilde {F}_{1}\left(\tilde
{L},\tilde {l},\tilde {G_{0}}\right)  ,
\end{equation}

To understand the effects of the argument $\theta_{3}$ and the
eccentricity $e_{2}$ of the outer planet of HD 82943 system, we
investigated the motion of phase space of HD 82943c according to
equation(\ref{eq6}) by using the same initial values for the
integration. In the analytical model,we remained
$\theta_{3}=0^{\circ}$, and let $e_{2}$ = 0.01,0.3 and 0.5
separately. Fig.3a exhibits the mean motion resonance argument
$\theta_{1}$ plotted against the semi-major axis $a$ of the inner
planet for the numerical results. Note that the small libration
with the amplitude of $\pm40^{\circ}$. Fig.3b shows the analytical
results given by the above coplanar model for the semi-major axis.
The estimated equilibrium point is about $(0.730, 0^{\circ})$. The
panel indicates that the 2:1 MMR could exist and be retained even
gradually changing the eccentricity of HD 82943b and implies that
the analytical results are quite consistent with the numerical
simulations in comparison with Fig3a. And we found the similar
results for $\theta_{3}=180^{\circ}$ by adopting the above values
of $e_{2}$.

For further experiments, we remained $e_{2}=0.1$ but varied
$\theta_{3}$,which was taken as $0^{\circ},20^{\circ},40^{\circ}$
and $120^{\circ},150^{\circ},180^{\circ}$,respectively. Fig.4a-4b
exhibit the computational results,in these figures we note that
for the same small eccentricity of the outer planet,the small
amplitude of libration of $\theta_{1}$ occurs when
$\theta_{3}=0^{\circ}$ and $180^{\circ}$,the large amplitude of
libration happens when $\theta_{3}=20^{\circ}$ and
$150^{\circ}$,and the circulation appears when
$\theta_{3}=40^{\circ}$ and $120^{\circ}$. Therefore, we can
safely conclude that the 2:1 MMR is easy to hold when
$\theta_{3}=0^{\circ},180^{\circ}$ for the small eccentricity of
the outer planet.

However,for some multi-planetary systems(such as GJ 876, HD
82943), $\theta_{3}$ is found to librate about $0^{\circ}$ or
$180^{\circ}$. From the previous analysis ,we comment that the
reason is that the special choice of $\theta_{3}$ is helpful to
preserve the mean motion resonance of two orbiting planets of such
systems.Furthermore,the orbits of the planets is well separated by
the mean motion resonance and then they are prevented from close
approaches free from strong perturbations of each other.At the end
of this section,we emphasize that this coplanar analytical model
for the semi-major axis is valid for the low eccentricity of the
outer planet(for No.1,No.17 and No.37,see Table 2). But we
surprisingly found the orbits of No.8 apparently cross during the
orbital evolution for the coplanar cases,and they are really in
the 2:1 MMR as mentioned. The following Section will explain
another mechanism for this peculiar case by holding their
eccentricities.

\subsubsection{Coplanar analytical model for the eccentricity}

The  difference of the apsidal longitudes $\theta_{3}$ is denoted
by
\begin{equation}
\label{eq7} \theta_{3} = \theta_{1} -
\theta_{2}=\tilde\omega_{1}-\tilde\omega_{2}
\end{equation}
\noindent where $\tilde\omega_{1},\tilde\omega_{2}$ are the
longitudes of the periastron respectively.For the case of No.8,
from Fig.5a,we notice that $\theta_{3}$ librates about
$180^{\circ}$,rather than $0^{\circ}$, with an amplitude of $\pm
30^{\circ}$,which indicates that the alignment  of the
pericenters(Chiang,Tabachnik \& Tremaine 2001;Lee \& Peale 2002)
for the HD 82943 system. The apsidal alignment corresponds to the
libration of the periastra longitudes of the two orbits, such that
the two planets have the same averaged rate of apsidal precession.
Recently, Malhotra(2002) described a dynamical mechanism for
establishing  apsidal alignment in a pair of planets that
initially move on nearly circular orbits by using classical
secular perturbation theory. Kinoshita \& Nakai(2002) developed a
semi-analytical secular method to explain the eccentricities of
the two planets of GJ 876 and suggested the stable mechanism of
the alignment of the pericenters. Now we again applied this method
to the HD 82943 system.

The Hamiltonian for the coplanar case is
\begin{equation}
\label{eq8}
F=F(a_1,a_2,e_1,e_2,\varpi_1,\varpi_2,\lambda_1,\lambda_2).
\end{equation}
Since we discuss the eccentricity of either of the planet in the
case of the 2:1 MMR, by eliminating short-periodic terms,the new
Hamiltonian reads:
\begin{equation}
\label{eq9} F^*=F^*(a_1,a_2,e_1,e_2,\theta_1,\theta_3),
\end{equation}
The degree of freedom of the new Hamiltonian(\ref{eq9}) is reduced
from four to two. However this Hamiltonian is not integrable. As
we discuss in Section 3.1.3, the semi-major axes of the two
resonant planets slightly change when they are in the 2:1 MMR, so
we can assume that $a_1$ and $a_2$ are constant for the critical
argument $\theta_1=0$. In final,the degree of freedom of the new
Hamiltonian is reduced to one:
\begin{equation}
\label{eq10} F^*=F^*(e_1,e_2,\theta_3),
\end{equation}
\noindent with conservation of the angular momentum:
\begin{equation}
\label{eq11}
\mu_{1}\sqrt{a_1(1-e_1^2)}+\mu_{2}\sqrt{a_2(1-e_2^2)}=const=H.
\end{equation}
\noindent We again eliminate $e_2$ in equation(\ref{eq10})
according to (\ref{eq11}) and we have
\begin{equation}
\label{eq12} F^*=F^*(e_1,\theta_3,H)
\end{equation}
\noindent Thus we can draw the level curves of the Hamiltonian and
understand the global behavior of $e_1$ and $\theta_3$. As the
eccentricity $e_1$ of the inner planet becomes large, we
numerically averaged the original Hamiltonian (\ref{eq8}) under
the condition of  the critical argument $\theta_1=0$ and the
angular momentum conservation (\ref{eq11}), and then obtain
numerically the averaged Hamiltonian (\ref{eq12}) with the
parameter of the angular momentum $H$. The technique of the
numerical averaging is described in the paper(Kinoshita \& Nakai
1985), in which the case of $\theta_1\neq 0$ is also discussed. We
draw the contour map of the Hamiltonian (\ref{eq12}) by taking
$\theta_3$ as the horizontal axis and $e_1$ or $e_2$ as the
vertical axis with the parameter $H$, which is determined from the
initial conditions. Figure 5b shows the contour map of the
Hamiltonian (\ref{eq12}). Because the numerical solution includes
the short-periodic terms, the numerically averaged solution is
shown in Figure 5b. This figure shows a good agreement between the
numerical solution and the semi-analytical secular solution. The
eccentricities of both planets are well restricted because of the
libration of $\theta_3$ around $180^{\circ}$. Here we should
mention that the period of the critical argument $\theta_1$ is
equal to that of the difference of the longitudes of the
periastron $\theta_3$, whose fact is the part of the stabilization
mechanism of for the case of No.8 orbit. Therefore,we can safely
conclude that the stability of the system of the coplanar crossing
orbits is simultaneously maintained by two mechanisms-the 2:1 MMR
and the apsidal alignment.


\subsection{HD 37124}
\subsubsection{Coplanar  system}
For the coplanar cases, we performed the numerical simulations of
100 pairs of  HD 37124 system  whose orbits were initially
generated in the neighborhood of the fitting results for the time
span of $10^{6}$ yr. For all the initial orbits, we remained the
semi-major axes of HD 37124b and HD 37124c unchanged. The
eccentricities are in the intervals of $[0.04,0.16]$ and
$[0.20,0.60]$,respectively. The arguments of periastra are
respectively in the span of
$[97^{\circ}-40^{\circ},97^{\circ}+40^{\circ}]$ and
$[265^{\circ}-120^{\circ},265^{\circ}+120^{\circ}]$. The nodal
longitudes and mean anomalies are made at random. However,we were
surprised to find that all of the coplanar configurations are
stable for $10^{6}$ yr,which indicates that the HD 37124 planetary
system is extremely steady for the above parameters. And the
simulation results again provide strong evidences for the fitting
results given by Butler(2002). In addition, we found that the
semi-major axis of HD 37124b remains 0.54 AU for most cases and
that of HD 37124c slightly librates about 2.95 AU(see Fig.6).
Fig.7 shows two typical changes of the eccentricities for the
stable orbits of HD 37124 system for $10^{6}$ yr:
(a)$e_{b}=0.1231$,$e_{c}=0.5592$,$\omega_{b}=64.36$,$\omega_{c}=309.56$.
The results of Fig.7a are similar to those of Butler(2002)(see
their Fig.14). The eccentricity of the inner planet undergoes
large variations ranging from 0.15 to 0.85,while that of the outer
planet varies from 0.15 to 0.60.
(b)$e_{b}=0.1313$,$e_{c}=0.4490$,$\omega_{b}=110.95$,$\omega_{c}=180.63$.
Fig.7b displays the small changes of the two eccentricities. We
should point out the range of the eccentricity of HD 37124c is
between 0 and 0.62 by analyzing the results of 100
integrations,which is in agreement with those of Butler(2002) who
stressed that  the system  will  be de-integrated for the
threshold initial eccentricity of the outer planet above 0.65. To
confirm this, several complemental integrations were done for
$e_{c}>0.65$. The outcomes show that the outer planet will escape
from the system in the time span less than $10^{6}$ yr and result
in the instability of the system.

\subsubsection{Mutual inclined orbit system}
For the non-coplanar cases, we only changed the inclination of HD
37124b $i_{b}$ and the other orbital parameters were unchanged. In
the initial conditions, the inclination of HD 37124c was always
taken as $0.5^{\circ}$ and  $i_{b}$ was respectively increased by
$10^{\circ}$,$20^{\circ}$,$30^{\circ}$,$40^{\circ}$, $50^{\circ}$,
$60^{\circ}$,$70^{\circ}$,$80^{\circ}$,$90^{\circ}$ with respect
to that of the outer planet. Thus the nine numerical experiments
were again carried out to understand the dynamical evolution for
the mutual inclined orbit system. From the results,we observed
that the semi-major axes of the two planets still retain well for
the integration of $10^{6}$ yr. Furthermore,we found the larger
amplitude variation of the eccentricity of HD 37124b(see
Fig.8),which can be explained by another mechanism-Kozai
resonance.

As Kozai(1962) pointed out, the resonance is present in the main
belt for the orbits of asteroids at very large inclinations,
because the perturbation of Jupiter will make the argument of
perihelion $\omega $ librate around 90$^{0}$ or 270$^{0}$. For new
Kozai hamiltonian, the transformed system is reduced to one degree
freedom and the Delaunay action variable $H = \sqrt {a(1-e^2)}\cos
i$ remains constant. The semi-major axis $a$ of the asteroid also
remains keeps unchanged during the secular orbital evolution after
eliminating short-period terms of the perturbation hamiltonian,
thus one could obtain $\bar {H} = \sqrt {\left( {1 - e^{2}}
\right)}\cos i = const$, so the inclination $i$ is minimum when
the eccentricity $e$ is maximum and $vice$ $versa$.


For our results,we found that this resonance takes place for three
cases of  $i_{b}=70.5^{\circ}$, $80.5^{\circ}$, $90.5^{\circ}$, in
the meanwhile $\omega$ undergoes the oscillations about
$90^{\circ}$ or $270^{\circ}$. Fig.8 displays that the argument of
periastron $\omega$ of HD 37124b librates about $90^{\circ}$ with
an amplitude of $\pm 76^{\circ}$ for the time span of $10^{6}$
yr,in fact,the phenomena also appears when we integrated the same
orbits for $10^{7}$ yr. Notice the coupled relationship-the
maximum eccentricity versus the minimum inclination and $vice$
$versa$. In addition,for this case we found the eccentricity and
the inclination of the outer planet are respectively in the ranges
$[0.47,0.57]$ and $[0^{\circ},40^{\circ}]$. Now we can use this
mechanism to explain why the eccentricities of the planets of some
extrasolar systems are so eccentric. Another important feature of
Kozai resonance is that the existence of the protection mechanism
means that it can keep the inner planet from close approaches. The
libration of $\omega $ around 90$^{0}$ or 270$^{0}$ means that at
periastron or apastron when the distance of the inner planet from
the parent star is closest to that of the outer planet, the inner
planet is also at a large distance from the fundamental plane.

Finally,we should underline that Kozai resonance is a very
effective mechanism to remain the stability of highly inclined
systems and may actually play an important part in the secular
dynamical evolution of the HD 37124 system,even for other
double-planet systems.

\section{Summary and Discussion}

In this paper, we have mainly explored the secular dynamical
evolution of the  HD 82943 and HD 37124 systems by simulating
various planetary configurations given in the neighborhood of the
fitting orbits. In final, we summarize some conclusions:

(1)For HD 82943 system, we carried out 100 integrations for the
time span of $10^{7}$ yr and finally found five stable
configurations. The results show that all of the stable orbits are
associated with the 2:1 MMR. Because of the small libration of the
resonant argument,the variations of the semi-major axes of the
planets can be greatly restricted to escape from close approaches.
On the other hand, by means of the analytical model for the
semi-major axis by considering the eccentricity of the outer
planet, we again discussed the motion of the phase space of the
inner planet for different $e_{2}$ and $\theta_{3}$, we detected
that the 2:1 MMR can be easily preserved when
$\theta_{3}=0^{\circ}$ or $180^{\circ}$ for relatively smaller
$e_{2}$. Still, the model for the semi-major axis is compared with
the numerical results and they are in good agreement with each
other,which both suggest that the existence of a 2:1 MMR can act
as one of the mechanisms to maintain the stability of the HD 82943
system.

The second mechanism is the alignment of the pericenters, and in
this paper we found that $\theta_{3}$ librates about $180^{\circ}$
for HD 82943 system in the simulations. In the meanwhile, the
eccentricities of the two planets undergo small oscillations for
the case,then the equi-Hamiltonian curves implies that the
numerical results are well consistent with the coplanar analytical
model for the eccentricities. In particular,we found the stable
orbits of No.8 apparently cross for the coplanar cases.
Generally,the apparent crossing orbits  become finally unstable
during secular orbital evolution. The numerical results suggest
that the two planets for No.8 are really in the 2:1 MMR. In our
opinion, we should point out that the stability of the new
discovery of the coplanar crossing orbits is maintained by both
2:1 MMR and the apsidal alignment.

(2)For HD 37124 system,we integrated 100 pair of the orbits of the
coplanar cases for $10^{6}$ yr and found that all configurations
are stable. However, for non-coplanar cases,we found that Kozai
resonance could frequently take place for the high inclination of
the inner planet with respect to the outer planet. The presence of
this resonance can explain the stable orbits of  the large
eccentricity of the inner planet of HD 37124 system and the small
libration of $\omega$ about 90$^{0}$ or 270$^{0}$ makes it
possible  the coupled relationship of the maximum eccentricity
with regard to the minimum inclination and the minimum
eccentricity  to the maximum inclination. We still believe that
Kozai mechanism could exist for other extrasolar planetary systems
of the mutual inclined configurations. And the relevant model will
be examined in our future work.

\acknowledgments

{We would thank Dr. Liao Xinhao for carefully reading the
manuscript and for constructive remarks and suggestions. This work
is financially supported by the National Natural Science
Foundations of China(Grant No.10173006),the Research Fund for the
Doctoral Program(RFDP) of Higher Education(Grant No.2000028416),
the Foundation of the Postdoctoral Science of China and the
Foundation of  Minor Planets of Purple Mountain Observatory.}

\clearpage

\begin{deluxetable}{lllll}
\tablewidth{0pt} \tablecaption{The parameters of HD 82943 and HD
37124 planetary systems} \tablehead{ \colhead{Parameter}    &
\colhead{HD 82943b} & \colhead{HD 82943c} & \colhead{HD 37124b} &
\colhead{HD 37124c}} \startdata
$M$sin$i$($M_{Jup}$)       & 1.63  & 0.88   &0.86   &1.01  \\
Orbital period $P$(days)   & 444.6 & 221.6  &153    &1942  \\
$a$(AU)                    & 1.16  & 0.73   &0.54   &2.95  \\
Eccentricity $e$           & 0.41  & 0.54   &0.10   &0.40  \\
$\Delta e$                 & 0.08  & 0.05   &0.06   &0.2   \\
$\omega$(deg)              & 117.8 & 138.0  &97     &265   \\
$\Delta\omega$(deg)        & 3.4   & 10.2   &40     &120  \\
\enddata
\end{deluxetable}

\clearpage
\begin{deluxetable}{ccccccc}
\tablewidth{0pt} \tablecaption{The orbital elements variations of
five stable orbits of  the two planets of HD 82943 system for the
time span of 10 M yr.} \tablehead{ \colhead{Run}    &
\colhead{$a_{\min } $(AU)} & \colhead{$a_{\max } $(AU)} &
\colhead{$e_{\min } $} & \colhead{$e_{\max } $} &
\colhead{$i_{\min }$(deg) } & \colhead{$i_{\max } $(deg)}  }
\startdata
1          &0.7221 &0.7418 &0.3223 &0.6932 &0.18   &31.76  \\
\nodata    &1.1460 &1.1779 &0.0005 &0.3473 &0.01   &12.22  \\
8          &0.7272 &0.7342 &0.3371 &0.5875 &0.13   &0.99   \\
\nodata    &1.1546 &1.1722 &0.4330 &0.5393 &0.00   &0.56   \\
17         &0.7238 &0.7391 &0.4613 &0.7392 &0.00   &1.55   \\
\nodata    &1.1475 &1.1729 &0.0351 &0.4154 &0.07   &0.88   \\
37         &0.7271 &0.7337 &0.2684 &0.6964 &0.05   &30.12  \\
\nodata    &1.1505 &1.1674 &0.0280 &0.3795 &0.02   &11.62  \\
92         &0.7264 &0.7339 &0.2905 &0.6046 &0.01   &13.44  \\
\nodata    &1.1547 &1.1740 &0.3736 &0.5164 &0.00   &5.67   \\

\enddata
\end{deluxetable}
\clearpage


\clearpage

\epsscale{1.0}
\begin{figure}
\figurenum{1}   \caption{This figure displays the variations of
the semimajor axis $a$, eccentricity and inclination of HD 82943b
and HD 82943c for the time span of 10 Myr for  No.17. Note that
the semi-major axes would undergo small oscillations for the whole
time span.\label{fig1} }
\end{figure}
\clearpage

\epsscale{1.00}
\begin{figure}
\figurenum{2}  \caption{(a) The left panel illustrates two
resonant arguments $\theta_{1}$ and $\theta_{2}$ vary with time
for  10 Myr. Notice that $\theta_{1}$ librates about
$180^{\circ}$,while $\theta_{2}$ librates about $0^{\circ}$.
(b)The right panel displays the semi-major axes of the inner and
outer planets of the system plotted against the resonant arguments
respectively. The equilibrium points are $(0.730,180^{\circ})$ and
$(1.165,0^{\circ})$,indicating the 2:1 MMR for the HD 82943
planetary system. \label{fig2} }
\end{figure}
\clearpage

\epsscale{1.00}
\begin{figure}
\figurenum{3}  \caption{(a)The left panel exhibits the mean motion
resonance argument $\theta_{1}$ plotted against the semi-major
axis $a$ of HD 82943c  for the numerical results. Note that the
small libration with the amplitude of $\pm40^{\circ}$.(b) The
analytical results are given by the coplanar model for the
semi-major axis. For $\theta_{3}=0^{\circ}$, $e_{2}$ = 0.01,0.3
and 0.5. The estimated equilibrium point is about $(0.730,
0^{\circ})$.The panel indicates that the 2:1 MMR could exist and
be retained even gradually changing the eccentricity of HD 82943b
and implies that the analytical results are quite consistent with
the numerical simulations in comparison with Panel (a).
\label{fig3}}
\end{figure}
\clearpage

\begin{figure}
\figurenum{4}  \caption{The phase space of the HD 82943c:(a)For
$e_{2}=0.1$,$\theta_{3} = 120^{\circ},150^{\circ},180^{\circ}$.
(b)For $e_{2}=0.1$,$\theta_{3} = 0^{\circ},20^{\circ},40^{\circ}$.
Note that for the same  $e_{2}$,the small amplitude of  libration
of $\theta_{1}$ occurs when $\theta_{3}=0^{\circ}$ and
$180^{\circ}$,the large amplitude of libration  for
$\theta_{3}=20^{\circ}$ and $150^{\circ}$,and the circulation for
$\theta_{3}=40^{\circ}$ and $120^{\circ}$. Panels (a) and (b)
reveal  the 2:1 MMR is easy to hold when
$\theta_{3}=0^{\circ},180^{\circ}$ for  smaller $e_{2}$.
\label{fig4} }
\end{figure}
\clearpage

\begin{figure}
\figurenum{5}  \caption{(a) The left panel displays that the
argument $\theta_{3}$ librates about $180^{\circ}$, with an
amplitude of $\pm 30^{\circ}$ for the HD 82943 system,which
implies a periastron alignment for this system.(b)The
equi-Hamiltonian curves for the eccentricities of the two planets
versus $\theta_{3}$.The thin lines are computed from numerically
averaged Hamiltonian by assuming that the planetary system is in
exact 2:1 MMR. The broad thick line in the libration region around
$\theta_{3}=180^{\circ}$ shows the solution obtained by numerical
integration, which suggests a good agreement with the
semi-analytical method for the case of the libration.
\label{fig5}}
\end{figure}
\clearpage

\begin{figure}
\figurenum{6}  \caption{ The variations of the semi-major axes of
HD 37124b and HD 37124c. Note that the semi-major axis of HD
37124b remained 0.54AU for $10^{6}$ yr,and that of HD 37124c
slightly librates about 2.95 AU. \label{fig6}}
\end{figure}
\clearpage

\begin{figure}
\figurenum{7}  \caption{Two typical variations of the
eccentricities for the stable orbits of  HD 37124 system for
$10^{6}$ yr:the broad line represents the changes of eccentricity
of HD 37124c,and the narrow line for those of HD
37124b.(a)$e_{b}=0.1231$,$e_{c}=0.5592$,$\omega_{b}=64.36$,$\omega_{c}=309.56$.The
results of Panel (a) are similar to those of Butler(2002)(see
their Fig.14). The eccentricity of the inner planet undergoes
large variations ranging from 0.15 to 0.85,while that of the outer
planet varies from 0.15 to 0.60.
(b)$e_{b}=0.1313$,$e_{c}=0.4490$,$\omega_{b}=110.95$,$\omega_{c}=180.63$.
Panel (b) shows the small changes of the two eccentricities. Note
that they separately librates about 0.20 and 0.50. \label{fig7}}
\end{figure}
\clearpage

\epsscale{0.80}
\begin{figure}
\figurenum{8} \caption{The case of Kozai resonance of HD 37124b
for mutual inclined orbits($i_{b}=70.50$). The argument of
periastron $\omega$ librates about $90^{\circ}$ an amplitude of
$\pm 76^{\circ}$ for the time span of $10^{6}$ yr . Note notice
the coupled relationship-the maximum eccentricity versus the
minimum inclination. As a matter of fact, the same results are
obtained even for  $10^{7}$ yr. \label{fig8}}
\end{figure}
\clearpage

\end{document}